\PassOptionsToPackage{hyphens}{url}

\documentclass[hidelinks]{juliacon} 
\usepackage{amsmath}
\allowdisplaybreaks

\usepackage{siunitx}

\usepackage{subcaption} 

\usepackage{tikz}

\usepackage{pgfplots}
\pgfplotsset{compat=1.16}

\usepackage{xspace}

\newcommand{\eg}[0]{{e.g.\@}\xspace}
\newcommand{\ie}[0]{{i.e.\@}\xspace}

\newcommand{\trixi}{Trixi.jl\xspace}

\begin{document}


\title{Adaptive numerical simulations with Trixi.jl: A case study of Julia for scientific computing}

\author[1]{Hendrik Ranocha}
\author[2, 3]{Michael Schlottke-Lakemper}
\author[4]{Andrew R. Winters}
\author[2]{Erik Faulhaber}
\author[5]{Jesse Chan}
\author[2, 3]{Gregor J. Gassner}
\affil[1]{Applied Mathematics, University of Münster, Germany}
\affil[2]{Department of Mathematics and Computer Science, University of Cologne, Germany}
\affil[3]{Center for Data and Simulation Science, University of Cologne, Germany}
\affil[4]{Department of Mathematics; Applied Mathematics, Linköping University, Sweden}
\affil[5]{Department of Computational and Applied Mathematics, Rice University, Houston, Texas, US}


\hypersetup{
pdftitle = {Adaptive numerical simulations with Trixi.jl: A case study of Julia for scientific computing},
pdfsubject = {JuliaCon 2019 Proceedings},
pdfauthor = {Hendrik Ranocha, Michael Schlottke-Lakemper, Andrew R. Winters, Erik Faulhaber, Jesse Chan, Gregor J. Gassner},
pdfkeywords = {Julia, Scientific Computing, Numerical Simulations, Conservation Laws, Discontinuous Galerkin Methods, Adaptive Mesh Refinement, Compressible Euler Equations, Ideal Magnetohydrodynamics, Entropy Stability, Shock Capturing},
}

\maketitle

\begin{abstract}
We present Trixi.jl, a Julia package for adaptive high-order numerical simulations
of hyperbolic partial differential equations. Utilizing Julia's strengths,
Trixi.jl is extensible, easy to use, and fast. We describe the main design choices
that enable these features and compare Trixi.jl with a mature open
source Fortran code that uses the same numerical methods.
We conclude with an assessment of Julia for simulation-focused scientific
computing, an area that is still dominated by traditional high-performance
computing languages such as C, C++, and Fortran.
\end{abstract}

\section{Introduction}

We are broadly interested in simulation-focused scientific computing, in particular
numerical approximations for hyperbolic partial differential equations (PDEs), computational fluid dynamics (CFD),
and related problems. The focus of our research ranges from specific applications
in CFD to general multi-physics coupling strategies to the development and analysis of the high-order numerical methods
our simulations are built upon. In addition, we are academics and involved in teaching
students in these areas of science. Thus, we would like to have a
code that is
\begin{enumerate}
  \item extensible for research and development,
  \item easy to understand and use for students and collaborators,
  \item fast enough for applied 3D problems.
\end{enumerate}
Additionally, we are greedy\footnote{See \url{https://julialang.org/blog/2012/02/why-we-created-julia} (accessed 2021-08-11)}
and wish to include all these features within a single code.
Roughly one year of collaborative work has resulted in the current version of
\trixi\footnote{\url{https://github.com/trixi-framework/Trixi.jl}}, providing adaptive
high-order numerical simulations of hyperbolic PDEs
in Julia \cite{bezanson2017julia}. Starting as an experiment, we have been able
to reach more and more of our goals with \trixi.

In this article, we present an overview of the main features and design decisions
of \trixi in Section~\ref{sec:design-of-trixi}, laying the ground for an extensible
and easy-to-use framework of high-order methods for hyperbolic PDEs. Next, we
compare the serial performance with a mature high-performance computing (HPC) Fortran code in
Section~\ref{sec:performance-comparison}, demonstrating that Julia is not generically
slower than traditional HPC languages (and can even be faster in this particular case).
Thereafter, we present an assessment of Julia for simulation-focused scientific
computing based on our experience with \trixi in Section~\ref{sec:assessment-of-julia}.
Finally, we summarize our findings and conclusions in Section~\ref{sec:summary}.

\section{Capabilities and design of Trixi.jl}
\label{sec:design-of-trixi}

\trixi is designed as a simulation framework and library of high-order methods for conservation laws
of the form
\begin{equation}
\label{eq:hcl}
  \partial_t u(t, x) + \sum_{j=1}^d \partial_{x_j} f^j(u) = s(t, x, u),
  \quad t \in (0, T), x \in \Omega,
\end{equation}
in $d \in \{1, 2, 3\}$ space dimensions. Here, the independent variables are
time $t$ and space coordinates $x \in \Omega \subset \mathbb{R}^d$. The conserved
quantities are denoted as $u$, \eg, mass, momentum, and energy for the compressible
Euler equations of an ideal gas.
The physical system is specified by the fluxes $f^j$ and the source term $s$.
In addition, suitable initial and boundary conditions (ICs, BCs) are required.
\trixi also handles non-conservative PDE terms as in the shallow water equations
or magnetohydrodynamics equations with divergence cleaning, where source terms
can depend on derivatives of~$u$.

\subsection{Main features of Trixi.jl}

As of version v0.3.55 (August 2021), \trixi concentrates mainly on discontinuous
Galerkin (DG) methods \cite{hesthaven2007nodal, kopriva2009implementing}.
In particular, it has a focus on entropy-conservative and -dissipative methods
\cite{tadmor1987numerical, lefloch2002fully, fisher2013high,
ranocha2018comparison, chen2017entropy}. Currently, \trixi offers the
following features:
\begin{itemize}
  \item 1D, 2D, and 3D simulations on line/quad/hex/simplex meshes
  \begin{itemize}
    \item Cartesian and curvilinear meshes
    \item Conforming and non-conforming meshes
    \item Structured and unstructured meshes
    \item Hierarchical quadtree/octree grid with adaptive refinement
    \item Forests of quadtrees/octrees with \texttt{p4est} \cite{burstedde2011p4est}
          via P4est.jl\footnote{\url{https://github.com/trixi-framework/P4est.jl}}
  \end{itemize}

  \item High-order matrix-free discontinuous Galerkin methods
  \begin{itemize}
    \item Kinetic energy preserving and entropy-stable methods
    \item Entropy-stable sub-cell shock capturing
    \item Sub-cell positivity-preserving limiting
  \end{itemize}

  \item Multiple governing equations
  \begin{itemize}
    \item Compressible Euler equations (optionally with self-gravity)
    \item Magnetohydrodynamics (MHD) equations
    \item Multicomponent compressible Euler and MHD equations
    \item Acoustic perturbation equations
    \item Hyperbolic diffusion for elliptic problems
    \item Lattice-Boltzmann equations (D2Q9 and D3Q27 schemes)
    \item Several scalar conservation laws (\eg, advection, Burgers)
  \end{itemize}

  \item Integration with the Julia package ecosystem and external tools
  \begin{itemize}
    \item Time integration methods from OrdinaryDiffEq.jl
    \item Automatic differentiation with ForwardDiff.jl
    \item In-situ visualization with Plots.jl
    \item Postprocessing with ParaView/Visit via Trixi2Vtk.jl\footnote{\url{https://github.com/trixi-framework/Trixi2Vtk.jl}}
  \end{itemize}
\end{itemize}

Currently, \trixi provides shared-memory parallelization via multithreading.
Initial parallelization with MPI is available for some mesh types,
but full support for distributed-memory parallelism is subject to ongoing work.

Some of the main features of \trixi are demonstrated in the following figures.
The detailed physical setups and numerical parameters as well as all code
necessary to reproduce these figures are available in the reproducibility
repository for this article \cite{ranocha2021adaptiveRepro}.

\begin{figure}[!h]
\vspace{-0.7em}
  \begin{subfigure}{0.52\linewidth}
    \includegraphics[width=\textwidth]{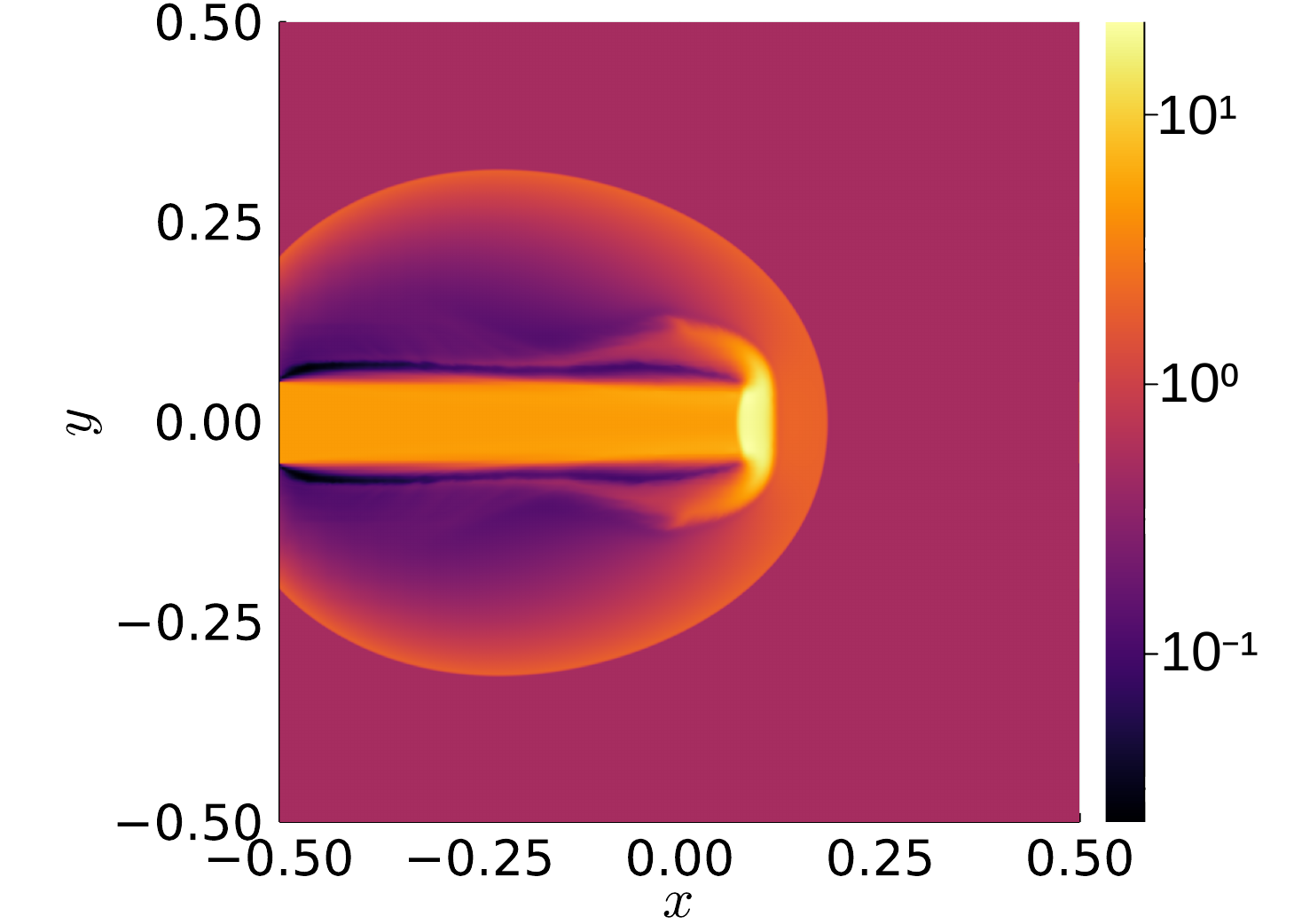}
    \caption{Density at time $t = 10^{-3}$.}
  \end{subfigure}%
  \hspace*{\fill}
  \begin{subfigure}{0.46\linewidth}
    \includegraphics[width=\textwidth]{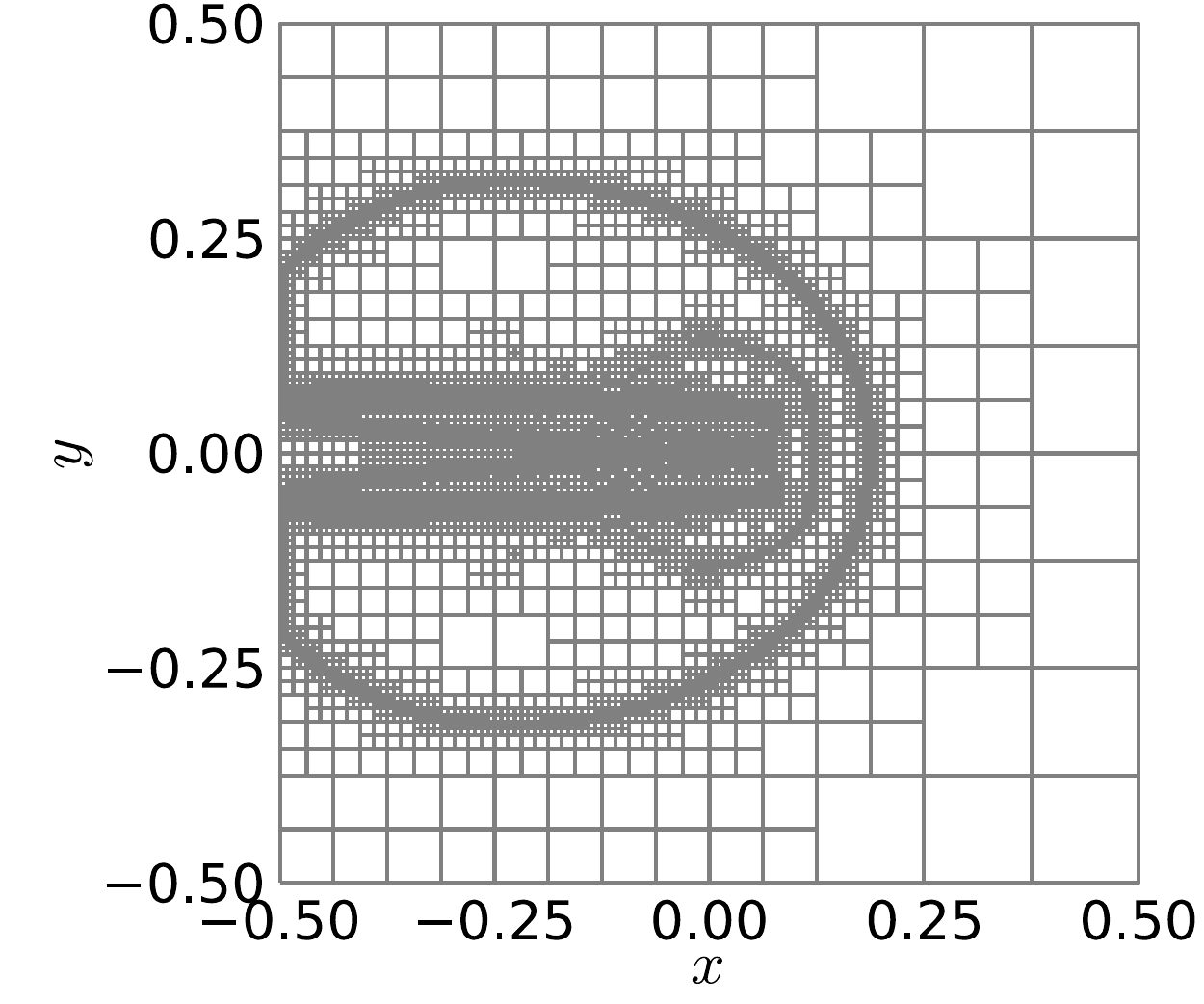}
    \caption{Mesh at time $t = 10^{-3}$.}
  \end{subfigure}%
  \caption{Numerical solutions of a supersonic jet with Mach number 2000 using
           sub-cell entropy-dissipative shock-capturing methods with sub-cell positivity-preserving
           limiters and adaptive mesh refinement for the compressible Euler equations.}
  \label{fig:jet}
\vspace{-0.7em}
\end{figure}

Figure~\ref{fig:jet} demonstrates sub-cell entropy-dissipative shock-capturing
methods with sub-cell positivity-preserving limiters and adaptive mesh refinement applied
to an astrophysical supersonic jet with Mach number 2000 \cite{liu2021oscillation}.

\begin{figure}[!ht]
  \begin{subfigure}{0.52\linewidth}
    \includegraphics[width=\textwidth]{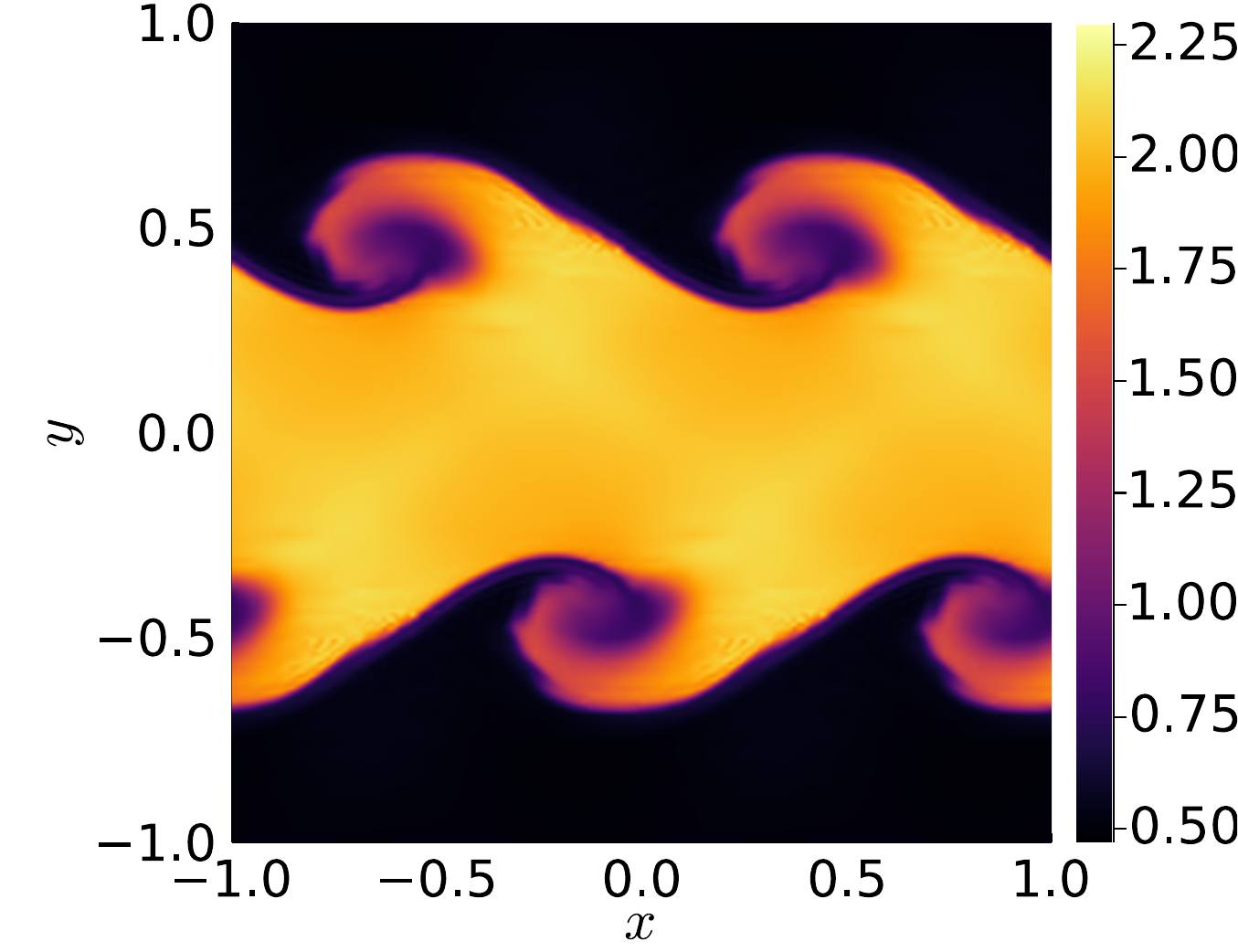}
    \caption{Density at time $t = 2$.}
  \end{subfigure}%
  \hspace*{\fill}
  \begin{subfigure}{0.46\linewidth}
    \includegraphics[width=\textwidth]{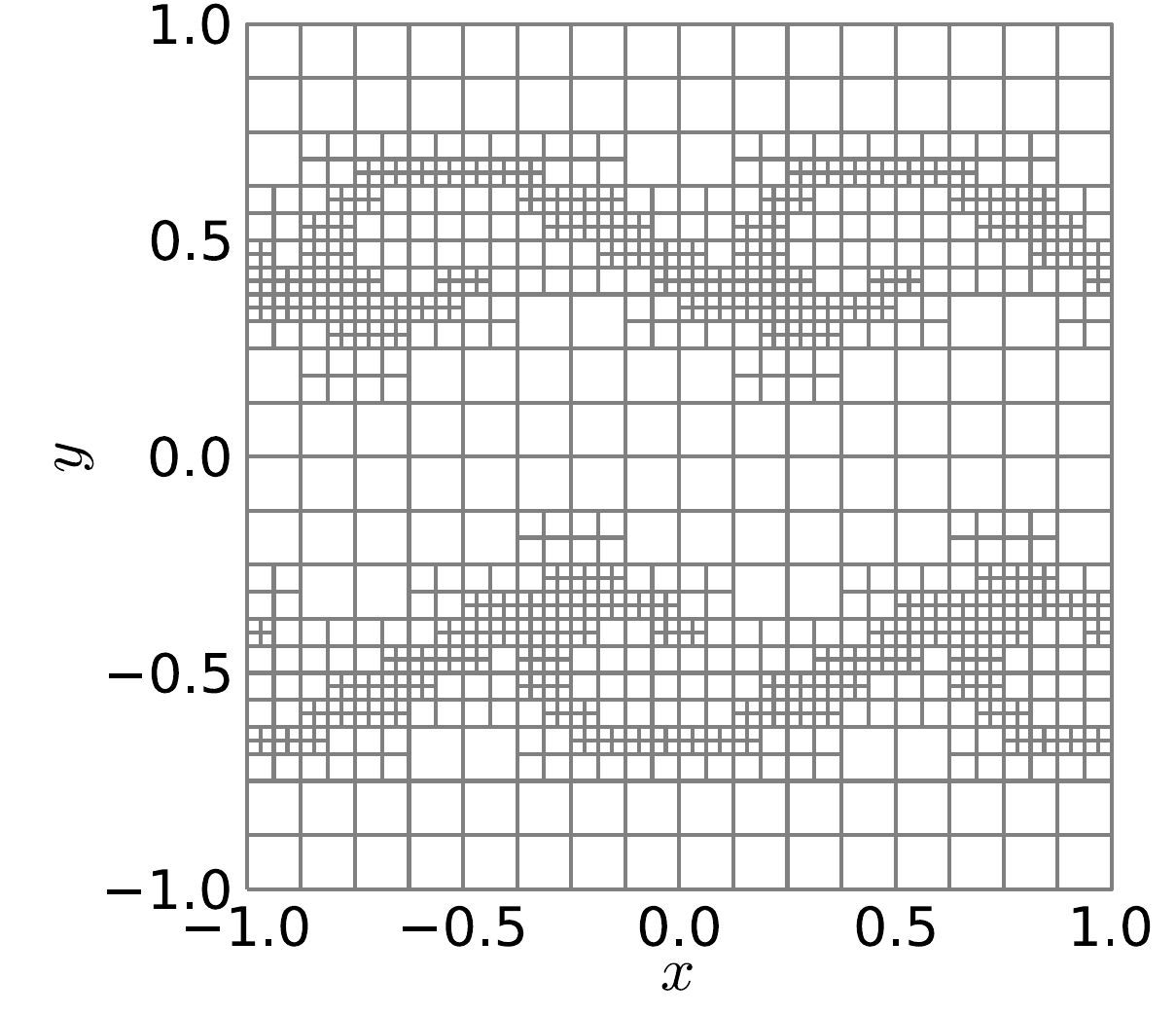}
    \caption{Mesh at time $t = 2$.}
  \end{subfigure}%
  \\
  \begin{subfigure}{0.52\linewidth}
    \includegraphics[width=\textwidth]{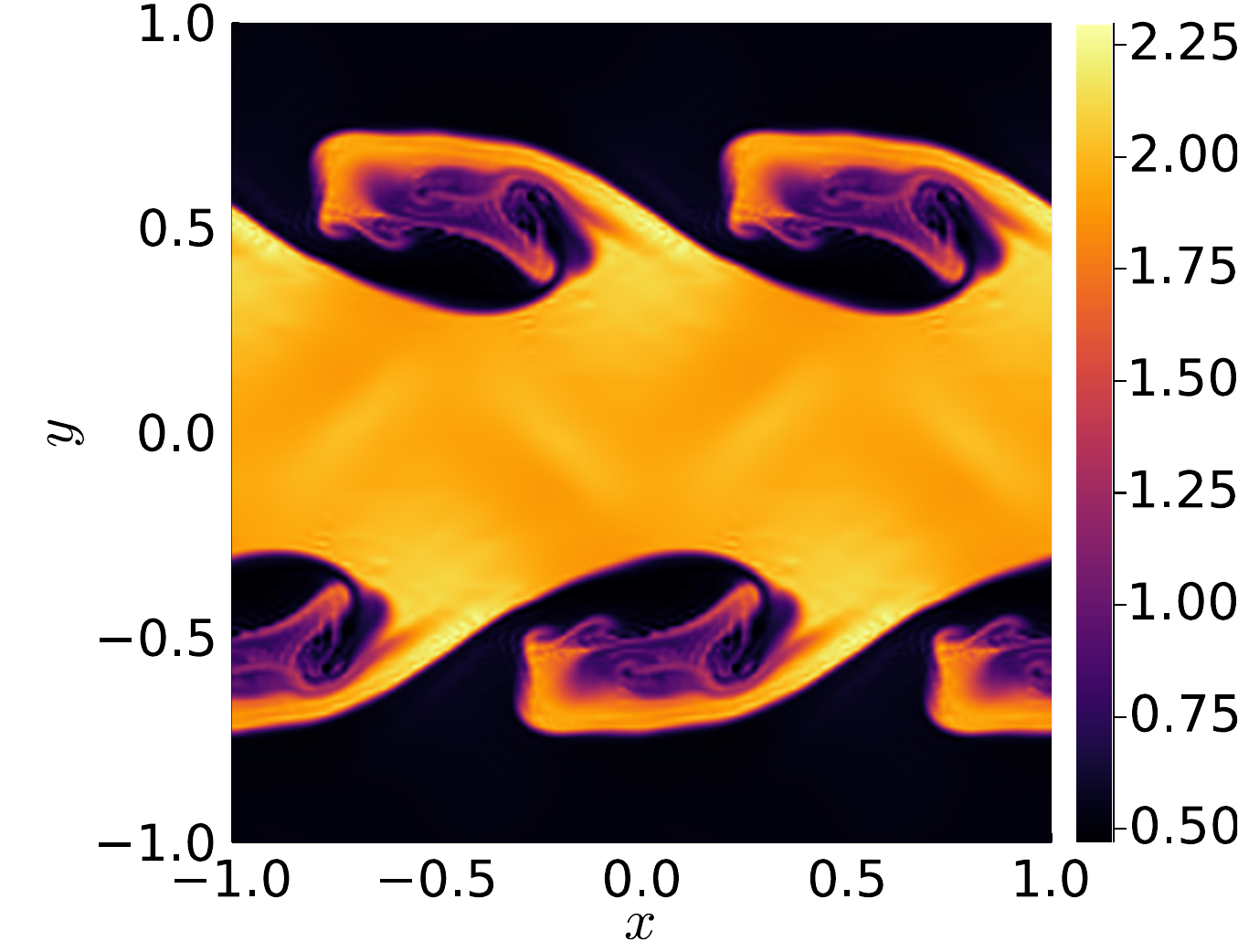}
    \caption{Density at time $t = 3$.}
  \end{subfigure}%
  \hspace*{\fill}
  \begin{subfigure}{0.46\linewidth}
    \includegraphics[width=\textwidth]{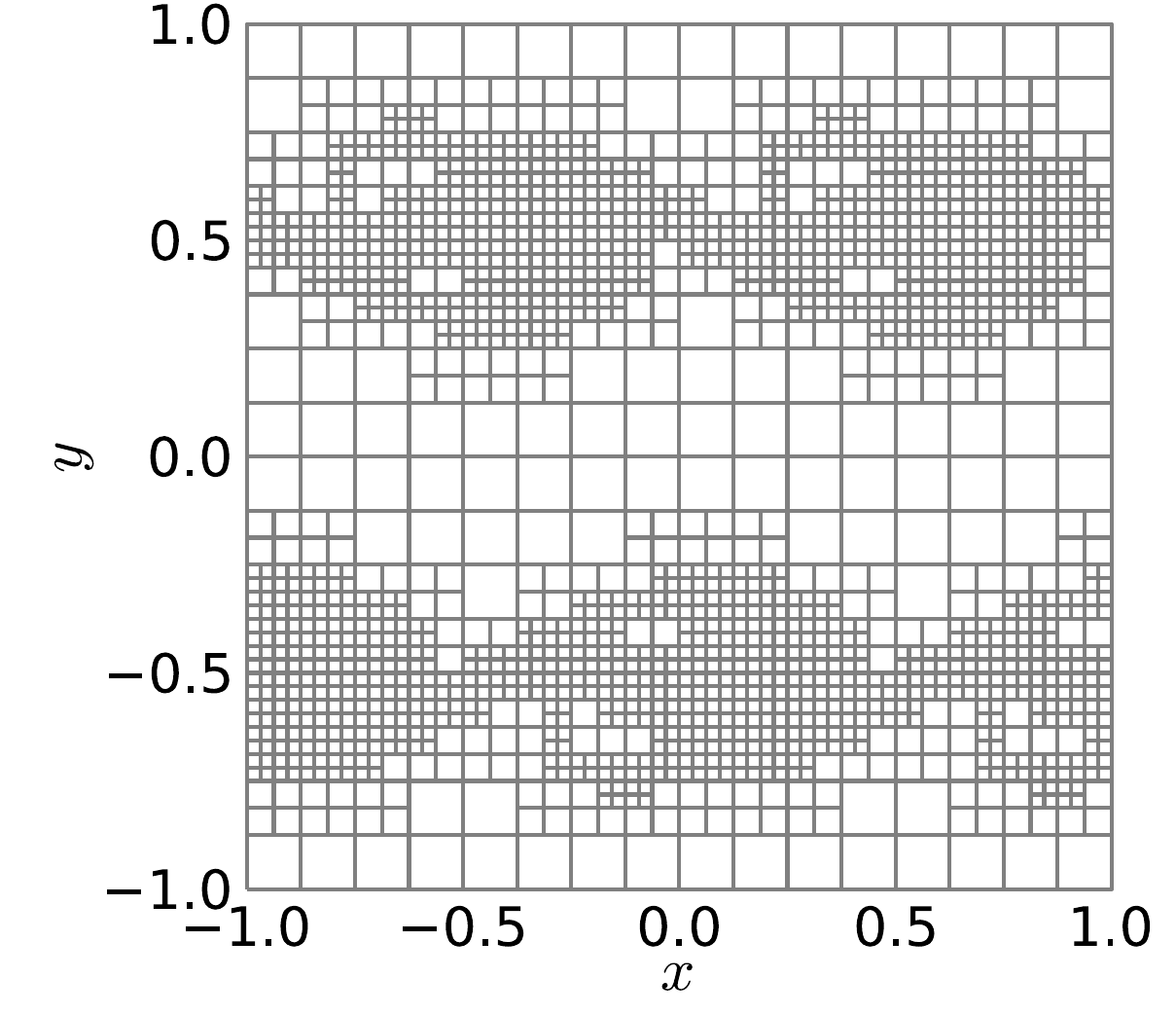}
    \caption{Mesh at time $t = 3$.}
  \end{subfigure}%
  \caption{Numerical solutions of a Kelvin-Helmholtz instability using
           entropy-stable methods and adaptive mesh refinement for the
           compressible Euler equations.}
  \label{fig:kelvin_helmholtzs}
\vspace{-0.7em}
\end{figure}

Figure~\ref{fig:kelvin_helmholtzs} demonstrates nonlinear stability
obtained with entropy-stable methods and adaptive mesh refinement applied to a
classical Kelvin-Helmholtz flow instability problem.

\begin{figure}[!ht]
\vspace{-0.7em}
  \begin{subfigure}{0.46\linewidth}
    \includegraphics[width=\textwidth]{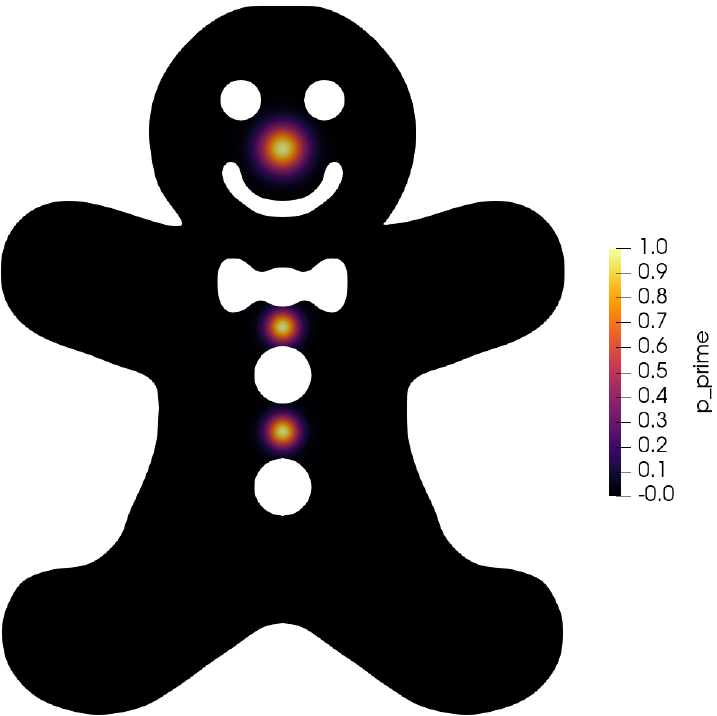}
    \caption{Acoustic pressure at time $t = 0$.}
  \end{subfigure}%
  \hspace*{\fill}
  \begin{subfigure}{0.46\linewidth}
    \includegraphics[width=\textwidth]{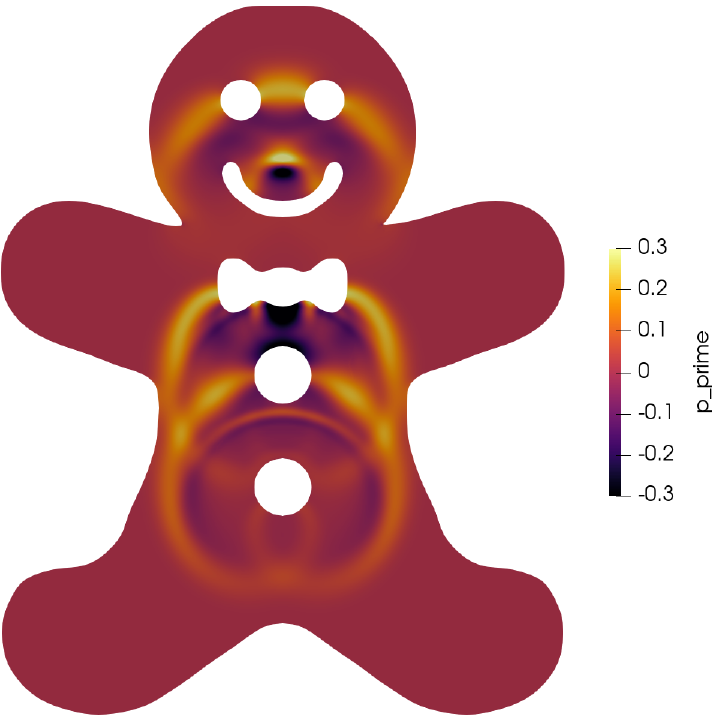}
    \caption{Acoustic pressure at time $t = 8$.}
  \end{subfigure}%
  \\
  \begin{subfigure}{0.46\linewidth}
    \includegraphics[width=\textwidth]{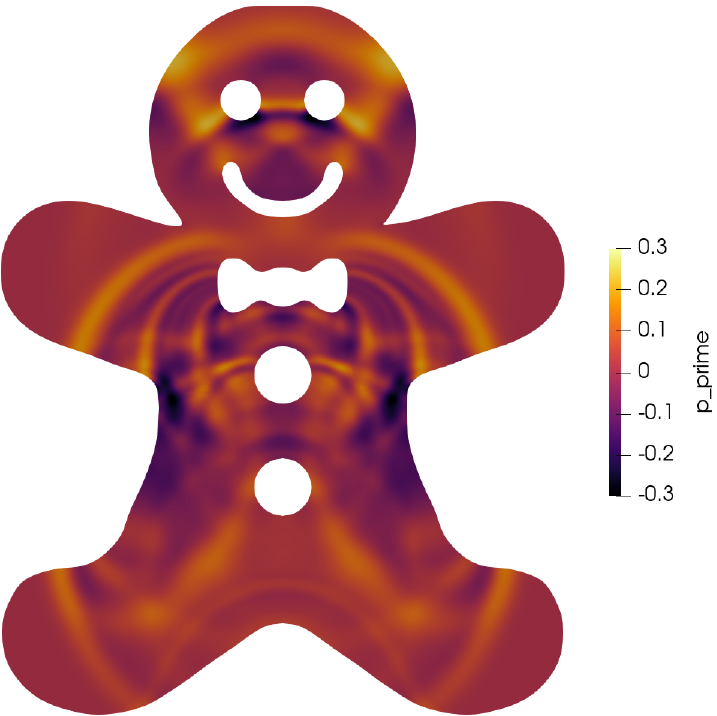}
    \caption{Acoustic pressure at time $t = 16$.}
  \end{subfigure}%
  \hspace*{\fill}
  \begin{subfigure}{0.46\linewidth}
    \includegraphics[width=\textwidth]{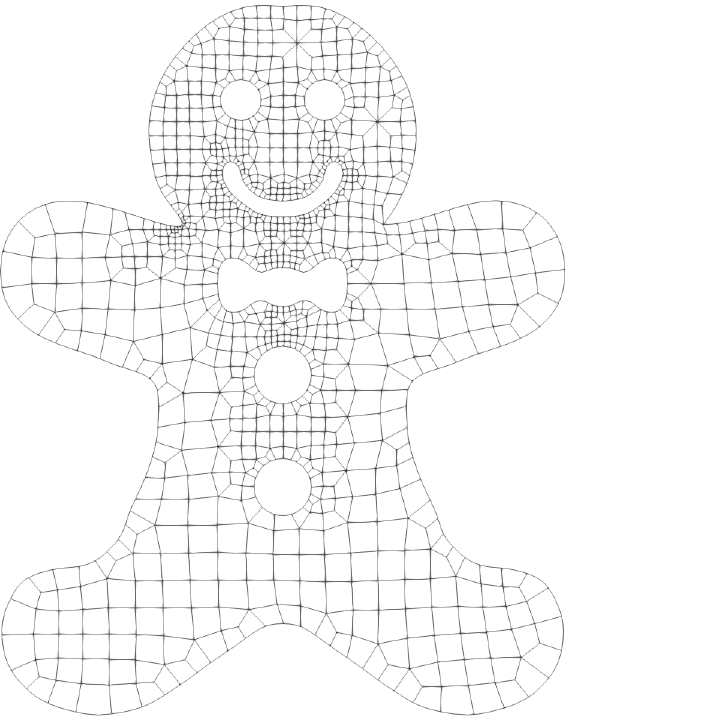}
    \caption{Unstructured quad mesh.}
    \label{fig:ginger_mesh}
  \end{subfigure}%
  \caption{Numerical solutions of pressure wave scattering for the acoustic
                perturbation equations at three points in time as well as
                the unstructured, curvilinear quadrilateral mesh.}
  \label{fig:pressure_waves}
\vspace{-0.7em}
\end{figure}

Figure~\ref{fig:pressure_waves} shows the approximation
of acoustic perturbation equations~\cite{ewert2003acoustic} wave scattering on a
curvilinear and unstructured domain. The 2D quadrilateral mesh used in the simulation
(Figure~\ref{fig:ginger_mesh}) was generated with
HOHQMesh.jl\xspace\footnote{\url{https://github.com/trixi-framework/HOHQMesh.jl}}.


\subsection{High-level overview of the code structure}

\trixi is built from the method of lines. Thus, a discretization of \eqref{eq:hcl}
is obtained in two steps. First, a spatial semidiscretization is created. Next,
the resulting ordinary differential equation (ODE) is solved using a time
integration method. Currently, \trixi focuses on the spatial semidiscretization
and uses mostly Runge-Kutta methods implemented in OrdinaryDiffEq.jl, which is part
of DifferentialEquations.jl \cite{rackauckas2017differentialequations}.

\begin{figure*}[th]
\centering
  \includegraphics[width=0.9\linewidth]{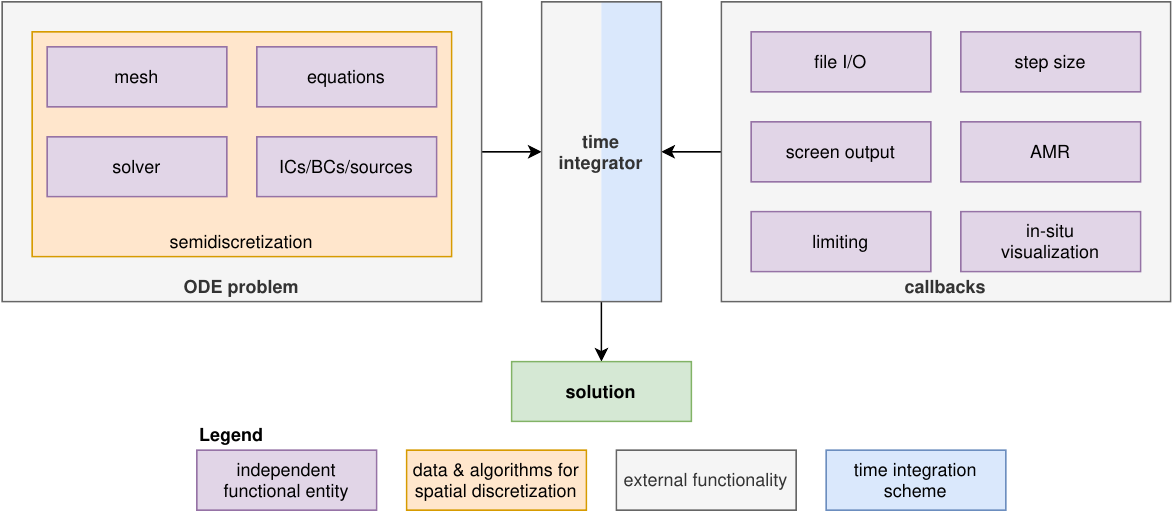}
  \caption{Schematic overview of the basic components in \trixi and how they
           interact.}
  \label{fig:trixi_global_overview}
\vspace{-0.7em}
\end{figure*}

Figure~\ref{fig:trixi_global_overview} presents an overview of the basic
components of \trixi. The most important structure is the semidiscretization,
which bundles all information about the spatial approximation. The mathematical-physical
model is determined by the \lstinline{equations}, the \lstinline{initial_condition},
\lstinline{boundary_conditions}, and possible \lstinline{source_terms}. The
\lstinline{solver} describes purely numerical parameters determining the specific
discretization method such as discontinuous Galerkin or finite difference
methods, kinetic energy preserving or shock-capturing approaches. Finally, the
\lstinline{mesh} has a necessarily hybrid role including information about the
spatial domain and its discretization.

Once information about the time span $[0, T]$ is provided, a semidiscretization
in \trixi can be converted into an ODE problem, which can be solved by methods
from OrdinaryDiffEq.jl. The flexible callback infrastructure of this ODE library
allows us to provide extended functionality for \trixi without modifying any
main loop. In particular, various tasks such as input/output operations, adaptive
mesh refinement (AMR), and positivity-preserving limiting are implemented using
callbacks. In case full control over the time step loop is required, e.g., to experiment with new
features that are not easily realized with the existing API, \trixi also implements its own time
integration schemes that mimic the interface of OrdinaryDiffEq.jl and that can be used as drop-in
replacements.

\subsection{Review of the main design choices}

The design of \trixi was guided by extensibility, ease of use, and efficiency
(in this particular order). Thus, it is likely possible to make some parts even
faster at the cost of simplicity and readability. Nevertheless, \trixi is already quite
fast, even compared to mature high-performance open source software,
as described in detail in Section~\ref{sec:performance-comparison}. In this
section, we focus mostly on the first two goals.

\subsubsection{Trixi.jl is a library}

The most important design decision was to create \trixi as a library. In the
context of simulation-focused scientific computing, there are many (open
or closed source) codes that are designed as monolithic applications that can
simulate a specific setup. These codes can often be configured to a limited extent
by specifying compile time options or editing parameter files. In contrast, users of
\trixi do not need to compile anything or learn parameter file syntax. Instead, they can
just load the package and write their simulation setups in Julia. Scripts that describe
a simulation setup for \trixi are referred to as ``elixirs''. As of version
v0.3.55 (August 2021), \trixi comes with approximately 200 example elixirs. A basic elixir
looks as follows:

\begin{lstlisting}[language = Julia]
# Load all required libraries
using Trixi, OrdinaryDiffEq, Plots

# Set up a 2D linear advection problem with
# advection velocity `a`
a = (1.0, 1.0)
equations = LinearScalarAdvectionEquation2D(a)

# Choose an initial condition coming with Trixi.jl
ic = initial_condition_convergence_test

# Set up a discontinuous Galerkin spectral element
# method with given polynomial degree and surface
# flux implemented in Trixi.jl
solver = DGSEM(polydeg=3,
               surface_flux=flux_lax_friedrichs)

# Create a uniformly refined mesh with periodic
# boundary conditions in a square domain
coordinates_min = (-1.0, -1.0) # lower left
coordinates_max = ( 1.0,  1.0) # upper right
mesh = TreeMesh(coordinates_min, coordinates_max,
                initial_refinement_level=4,
                n_cells_max=10^5, periodicity=true)

# Create semidiscretization with all spatial
# discretization-related components
semi = SemidiscretizationHyperbolic(
    mesh, equations, ic, solver)

# Create an ODE problem from the semidiscretization
# with time span from 0.0 to 1.0
ode = semidiscretize(semi, (0.0, 1.0))

# Evolve the ODE problem in time using `solve` from
# OrdinaryDiffEq with adaptive time stepping
sol = solve(ode, RDPK3SpFSAL49(), abstol=1.0e-6,
    reltol=1.0e-6, save_everystep=false)

# Plot the numerical solution at the final time
plot(sol)
\end{lstlisting}

\subsubsection{Functions are pure Julia functions}

Since functions are first-class citizens in Julia, they can be passed around
and used efficiently. Thus, everything that acts like a function can be used
in \trixi, \eg, to define initial/boundary conditions. While the flexibility
to change initial/boundary conditions
is quite common, two-point numerical fluxes are also just functions with a
specified signature. Thus, users can implement numerical fluxes in their
own code and they will work and be as efficient as if they were implemented
directly in \trixi.

\subsubsection{Ingredients use common interfaces and can be exchanged}

Abstractions such as (variants of) the \lstinline{solver}, the \lstinline{mesh},
and the \lstinline{equations} use a common interface. Utilizing multiple
dispatch in Julia, internal implementations are specialized accordingly. For
example, changing the volume terms of the DG discretization from the standard
weak form to entropy-stable and/or shock-capturing methods can be achieved by
passing one additional parameter to the \lstinline{DGSEM} constructor in the
example above. Moreover, there is no hidden global state. This means that multiple instances
of similar structures can be instantiated simultaneously. In particular, multiple
semidiscretizations (in possibly different spatial dimensions) can be created
and used in the same code.

\subsubsection{New physics can be specified with minimal effort}

Instead of providing only the very basic discretization ingredients, as with
other open source libraries for PDEs, \trixi also includes some widely used
physical systems as well as analysis routines, such as computation of the integrated kinetic energy
for the compressible Euler equations,
to make it easy to use out of the box. Nevertheless, users are not restricted
to the physics models bundled in \trixi. To set up a new type of \lstinline{equations},
it is only necessary to create an appropriate \lstinline{struct} containing all
parameters and to implement pointwise operations such as the calculation of
the fluxes $f^j$ in \eqref{eq:hcl} or two-point numerical fluxes. Due to the Julia language
design with just-ahead-of-time compilation, these fluxes can be inlined into the
library functions of \trixi. Thus, the physics is completely separated from
the \lstinline{solver} but remains computationally efficient. In particular, this allows
a user to reuse the same numerical fluxes for discontinuous Galerkin methods, finite
difference methods, and variants of finite volume methods.
Due to favoring
simplicity over excessively generic code, it remains comparatively straightforward to modify
existing \trixi implementations if a new feature requires modifications to methods.

\subsubsection{There is no spooky action at a distance}

In monolithic code bases, it is often necessary to implement a new
feature for all combinations of possible (compile time) options. This effort
can make it difficult to experiment with new ideas. In contrast, Julia's dynamic
nature, multiple dispatch, and just-ahead-of-time compilation allow us to implement only those features
that are strictly necessary. For example, if a user wants to simulate a new
physics model only on Cartesian grids with smooth solutions, there is no need
to implement anything for curvilinear coordinates or shock-capturing approaches.
In addition, it allows a user to extend \trixi step-by-step with new capabilities.
For example, there are ongoing efforts to incorporate summation-by-parts finite
difference methods via SummationByPartsOperators.jl \cite{ranocha2021sbp} and
discontinuous Galerkin methods on simplex elements via StartUpDG.jl\footnote{\url{https://github.com/jlchan/StartUpDG.jl}} in \trixi.
This is feasible because the modular design of \trixi gives users the flexibility to pick
a subset of the available meshes/methods, selected via multiple dispatch,
to test their newly implemented features.

\section{Performance comparison with Fortran}
\label{sec:performance-comparison}

FLUXO\footnote{\url{https://gitlab.com/project-fluxo/fluxo}} is an open source
Fortran code implementing discontinuous Galerkin methods on unstructured hexahedral
meshes in 3D for advection-diffusion equations. It provides
the same kind of modern, flux differencing DG methods
to achieve entropy conservation/dissipation or kinetic energy preservation that are used in \trixi.
At the same time, both codes have capabilities the other does not, \eg,
support for parabolic equations and multi-node parallelism in FLUXO,
or multiple mesh types and more physics setups in \trixi. Nevertheless,
both \trixi and FLUXO share a common set of features, \ie, high-order DG
methods on 3D curvilinear meshes, which allows a reasonable comparison of their performance.

\subsection{Description of the setup}

Here, we compare the serial performance of \trixi and FLUXO when solving a
hyperbolic PDE in three space dimensions on curvilinear hexahedral meshes.

For the problem setup we consider a periodic box of the domain $[-1,1]^3$ with four elements in each
spatial direction. This results in $64 (\text{\lstinline{polydeg}}+1)^3$ degrees
of freedom (DG nodes) per equation when polynomials of degree \lstinline{polydeg} are used.
To make the mesh curvilinear, the interior of the box is heavily warped by
a mapping adapted from \cite{chan2019efficient}.
Specifically, we map Cartesian reference coordinates $(\xi, \eta, \zeta) \in [-1, 1]^3$
to physical coordinates $(x, y, z)$ via the transformation
\begin{equation}
\begin{aligned}
  y &= \eta  + 0.15 \bigl( \cos(1.5 \pi \xi) \cos(0.5 \pi \eta) \cos(0.5 \pi \zeta) \bigr),
  \\
  x &= \xi   + 0.15 \bigl( \cos(0.5 \pi \xi) \cos(2 \pi y) \cos(0.5 \pi \zeta) \bigr),
  \\
  z &= \zeta + 0.15 \bigl( \cos(0.5 \pi x) \cos(\pi y) \cos(0.5 \pi \zeta) \bigr).
\end{aligned}
\end{equation}
To integrate up to the
final time $T=1.0$, both codes use the five-stage, fourth-order low-storage explicit
Runge-Kutta method of Carpenter and Kennedy~\cite{CarpenterKennedy1994}.
A stable explicit time step is adaptively computed according to the local maximum wave
speed, the relative grid size, and an adjustable Courant-Friedrichs-Lewy (CFL)
coefficient $\text{CFL}\in(0,1]$~\cite{gassner2011}. For the performance computations presented herein
we fix this coefficient to be $\text{CFL}=0.5$.

\begin{figure}[!h]
\centering
  \includegraphics[width=0.9\linewidth]{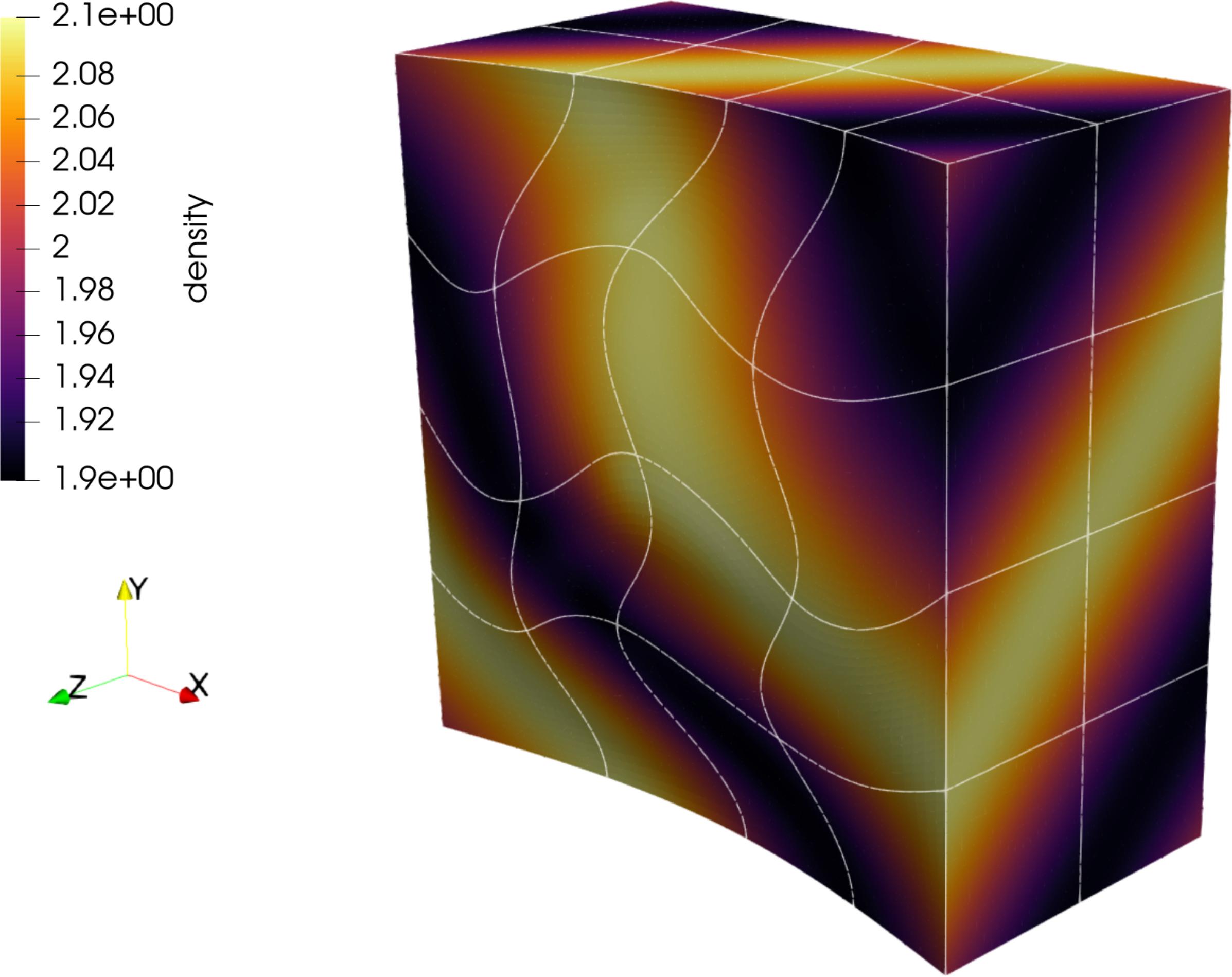}
  \caption{Initial density of the compressible Euler problem on a slice of the
           periodic, curved mesh used for the performance benchmarks.}
  \label{fig:pid_euler_mesh}
\end{figure}

We compare the performance of FLUXO and \trixi for two smooth nonlinear problems,
a manufactured solution for the 3D compressible Euler equations and Alfv\'{e}n wave propagation for
the 3D ideal magnetohydrodynamics (MHD) equations~\cite{gassner2009,altmann2012}. These initial conditions, available in FLUXO and
\trixi, are typically used to demonstrate the high-order accuracy and convergence properties
of the frameworks.
The curved mesh and the initial density for the compressible Euler problem are
visualized in Figure~\ref{fig:pid_euler_mesh}.

For these comparisons we examine the performance of FLUXO and \trixi for the weak
form as well as flux differencing implementations of the DG solver. For the weak form simulations we compute
the coupling between elements with a Harten-Lax-van Leer (HLL) numerical surface flux function~\cite{Toro2009,li2005hllc}.
For the flux differencing solver, which introduces an additional numerical volume flux function, we use an entropy-conservative flux
function in both the volume and at the surface. For the compressible Euler equations
this flux is also kinetic energy preserving~\cite{ranocha2020icosahom, ranocha2018thesis}
while for the ideal MHD equations it is kinetic and magnetic energy preserving~\cite{hindenlang2019}. We note that both ideal MHD implementations
require additional computational effort due to non-conservative terms necessary for the numerical approximation
to remain entropy consistent~\cite{bohm2020}. Additionally, the weak form DG solver type is recovered if one uses a central flux
in the volume of the flux differencing DG solver, albeit with additional computational cost~\cite{gassner2016split}.

All simulations for the performance comparison were run on the HPC system Tetralith
provided by the Swedish National Infrastructure for Computing (SNIC).
Each node of Tetralith has two sockets, each with an Intel(R) Xeon(R) Gold 6130 @ 2.10GHz with 96GiB of memory.
FLUXO was compiled using the Intel compiler suite v18.0.3 and \trixi used Julia v1.6.1.
The serial performance results were obtained on a single core.
We ran each configuration of a particular physical setup, polynomial degree, and solver type
for each code five times, taking the smallest value from each for the comparison results.
We ran the same PID tests in FLUXO compiled with GCC v6.4.0 but found that
the version compiled using Intel was between \SI{5}{\percent} to \SI{23}{\percent} faster.
Thus, only results from the Intel compiler are included.

To assess the performance of the two codes, we vary the polynomial degree
(\lstinline{polydeg}) from 3 to 15.
This corresponds to considering DG approximations of increasing spatial accuracy, from fourth up to sixteenth order.
The metric we use to analyze and compare FLUXO and \trixi simulations is a performance index (PID) that
measures the time required to advance a single degree of freedom (DOF) from one stage of the explicit Runge-Kutta
time integration scheme to the next. It is computed as
\begin{equation}
\label{eqn:pid}
\text{PID} = \frac{\text{wall-clock time}}{\#\text{time steps}\cdot 5\cdot\#\text{elements}\cdot(\text{\lstinline{polydeg}}+1)^3},
\end{equation}
where $5$ is the number of Runge-Kutta stages per time step for the selected time-stepping method. In other words, the PID
measures the run time required for the right hand side (RHS) evaluation of each DOF.

The code and detailed information necessary to reproduce these numerical
experiments are available in the accompanying repository \cite{ranocha2021adaptiveRepro}.

\subsection{Results of the performance comparison}

\begin{figure}[!ht]
\centering
  \begin{subfigure}{\linewidth}
    \begin{tikzpicture}[
      font=\footnotesize
      ]
      \begin{axis}[
          xlabel={Polynomial degree},
          ylabel={Time/RHS/DOF {[sec]}},
          width=\textwidth,
          height=0.66\textwidth,
          legend columns=-1,
          legend pos=north west,
          ymin=0.0,
          ymax=2.0e-6,
          grid=major,
        ]
        \addplot[blue, mark=*, solid] 
          table [x index=0, y index=2]{pid_data/pids_mhd_weak.dat};
          \addlegendentry{FLUXO}
        \addplot[red, mark=diamond*, solid] 
          table [x index=0, y index=1]{pid_data/pids_mhd_weak.dat};
          \addlegendentry{\trixi}
        \node [rotate=12] at (axis cs:13.5,5.75e-7) {Central flux};
        \draw (axis cs:13,6.0e-7) -- (axis cs:12.4,7.1e-7);
        \draw (axis cs:13.4,6.3e-7) -- (axis cs:13.7,1.075e-6);

        \addplot[blue, mark=*, dashed] 
          table [x index=0, y index=2]{pid_data/pids_mhd_flogor.dat};
        \addplot[red, mark=diamond*, dashed] 
          table [x index=0, y index=1]{pid_data/pids_mhd_flogor.dat};
        \node [rotate=20] at (axis cs:7,1.2e-6) {EC flux};
        \draw (axis cs:8.4,1.1e-6) -- (axis cs:7.75,1.15e-6);
        \draw (axis cs:7,1.1e-6) -- (axis cs:7.5,6.5e-7);
      \end{axis}
    \end{tikzpicture}%
    \caption{Absolute run times.}
  \end{subfigure}%
  \\
  \begin{subfigure}{\linewidth}
    \begin{tikzpicture}[
      font=\footnotesize
      ]
      \begin{axis}[
          xlabel={Polynomial degree},
          ylabel={Time relative to FLUXO {[-]}},
          width=\textwidth,
          height=0.66\textwidth,
          legend columns=-1,
          legend pos=south west,
          ymin=0.0,
          ymax=1.1,
          ytick={0, 0.2, 0.5, 0.6, 0.7, 0.8, 0.9, 1},
          grid=major,
        ]
        \addplot[blue, mark=*, solid, forget plot] 
          table [x index=0, y index=2]{pid_data/pids_mhd_weak_relative.dat};
        \addplot[red, mark=diamond*, solid] 
          table [x index=0, y index=1]{pid_data/pids_mhd_weak_relative.dat};
          \addlegendentry{Central flux}

        \addplot[blue, mark=*, dashed, forget plot] 
          table [x index=0, y index=2]{pid_data/pids_mhd_flogor_relative.dat};
        \addplot[red, mark=diamond*, dashed] 
          table [x index=0, y index=1]{pid_data/pids_mhd_flogor_relative.dat};
          \addlegendentry{EC flux}
      \end{axis}
    \end{tikzpicture}
    \caption{Run time relative to FLUXO.}
  \end{subfigure}%
  \caption{Run time per right-hand side evaluation and degree of freedom for
           the flux differencing DG discretization of the 3D ideal MHD equations.
           Two configurations are compared using either the central volume flux
           (algebraically equivalent to the weak form DG solver) and the entropy
           conservative (EC) volume flux.}
  \label{fig:PID-MHD}
\end{figure}
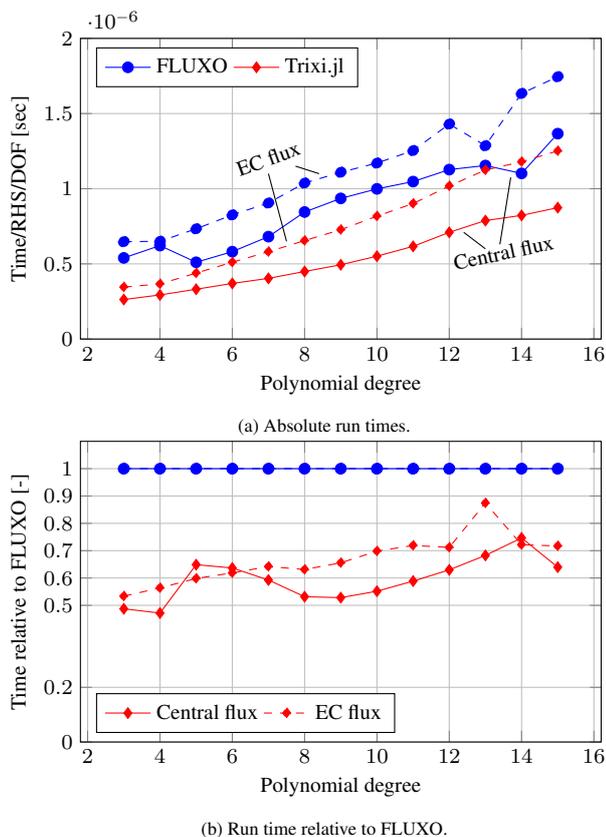

The results of these performance measurements are visualized for the 3D compressible Euler equations
in Figure~\ref{fig:PID-Euler} and for the 3D ideal MHD equations in Figure~\ref{fig:PID-MHD}.
We present the absolute PID timings for both codes in the top portion of these two figures and
a relative comparison using FLUXO as reference in the bottom portion; smaller values thus are always
better.

\begin{figure}[!ht]
\centering
  \begin{subfigure}{\linewidth}
    \begin{tikzpicture}[
      font=\footnotesize
      ]
      \begin{axis}[
          xlabel={Polynomial degree},
          ylabel={Time/RHS/DOF {[sec]}},
          width=\textwidth,
          height=0.66\textwidth,
          legend columns=-1,
          legend pos=north west,
          ymin=0.0,
          ymax=2.0e-6,
          grid=major,
        ]
        \addplot[blue, mark=*, solid] 
          table [x index=0, y index=2]{pid_data/pids_euler_weak.dat};
          \addlegendentry{FLUXO}
        \addplot[red, mark=diamond*, solid] 
          table [x index=0, y index=1]{pid_data/pids_euler_weak.dat};
          \addlegendentry{\trixi}
        \node at (axis cs:13,3.5e-7) {Weak form};

        \addplot[blue, mark=*, dashed] 
          table [x index=0, y index=2]{pid_data/pids_euler_ranocha.dat};
        \addplot[red, mark=diamond*, dashed] 
          table [x index=0, y index=1]{pid_data/pids_euler_ranocha.dat};
        \node [rotate=15] at (axis cs:9,8.2e-7) {Flux differencing};
        \draw (axis cs:9,9.2e-7) -- (axis cs:8.5,10.2e-7);
        \draw (axis cs:9,7.2e-7) -- (axis cs:8.5,5.2e-7);
      \end{axis}
    \end{tikzpicture}%
    \caption{Absolute run times.}
  \end{subfigure}%
  \\
  \begin{subfigure}{\linewidth}
    \begin{tikzpicture}[
      font=\footnotesize
      ]
      \begin{axis}[
          xlabel={Polynomial degree},
          ylabel={Time relative to FLUXO {[-]}},
          width=\textwidth,
          height=0.66\textwidth,
          legend columns=-1,
          legend pos=south west,
          ymin=0.0,
          ymax=1.1,
          ytick={0, 0.2, 0.4, 0.5, 0.6, 0.8, 1},
          grid=major,
        ]
        \addplot[blue, mark=*, solid, forget plot] 
          table [x index=0, y index=2]{pid_data/pids_euler_weak_relative.dat};
        \addplot[red, mark=diamond*, solid] 
          table [x index=0, y index=1]{pid_data/pids_euler_weak_relative.dat};
          \addlegendentry{Weak form}

        \addplot[blue, mark=*, dashed, forget plot] 
          table [x index=0, y index=2]{pid_data/pids_euler_ranocha_relative.dat};
        \addplot[red, mark=diamond*, dashed] 
          table [x index=0, y index=1]{pid_data/pids_euler_ranocha_relative.dat};
          \addlegendentry{Flux differencing}
      \end{axis}
    \end{tikzpicture}
    \caption{Run time relative to FLUXO.}
  \end{subfigure}%
  \caption{Run time per right-hand side evaluation and degree of freedom for
           different DG discretizations of the 3D compressible Euler equations
           in FLUXO (Fortran) and Trixi.jl (Julia).}
  \label{fig:PID-Euler}
\end{figure}
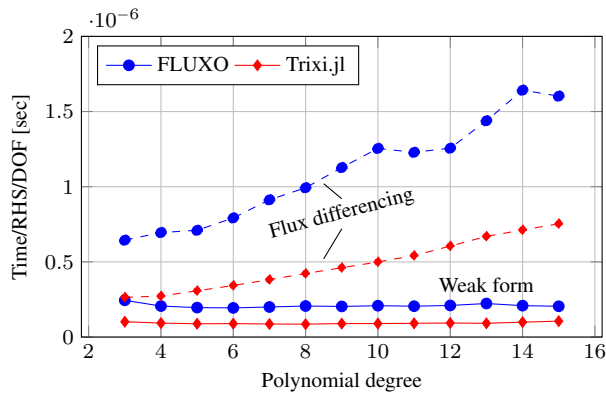
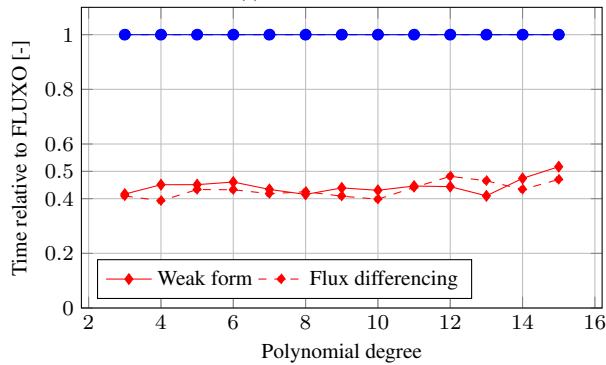

From these results, it is clear that \trixi is more than 2x faster for the compressible Euler equations
and more than 1.5x faster for the ideal MHD equations than the Fortran code FLUXO.
It is also possible to see the additional computational effort for the entropy-conservative flux
differencing DG methods for both physical systems. This is because the numerical fluxes are more
expensive in
terms of computational cost, and require special care to optimize their performance~\cite{hendrik_blog,ranocha2021efficient}.
We reiterate that both codes implement the same numerical nodal DG methods on curvilinear meshes used in these tests.
This demonstrates the suitability of Julia for this kind of simulation-focused scientific computing.

While our Julia code is faster than the mature HPC Fortran code FLUXO for this non-trivial
example, we do not claim that Julia is generally faster than Fortran, C, or C++.
Instead, we would like to emphasize that well-written Julia code can be \emph{at least
as fast} as code written in these traditional scientific computing languages,
as demonstrated also by several microbenchmarks \cite{bezanson2017julia}.
Further, we expect to be able to achieve similar performance with either language
by spending enough time and effort to optimize the respective codes.
\trixi owes its performance optimizations in part to the code introspection and
profiling tools available in Julia. Similar tools used to optimize the performance in other
languages are often not as easy to use as their Julia counterparts. Therefore,
we developed some additional performance improvements in \trixi and ported them
to FLUXO later \cite{ranocha2021efficient}. Moreover, \trixi uses slightly
different implementations of computationally intensive kernels and stores marginally
less information in memory, recomputing some terms instead. Additionally, the
manufactured solution of the compressible Euler equations uses source terms
containing sine and cosine terms, which are computed together by \lstinline{sincos}
in \trixi but require individual calls to \lstinline{sin} and \lstinline{cos} in
FLUXO.

Finally, the different mesh types available in \trixi allow further optimizations.
For example, if the computational domain is essentially a cube (or square, line)
without need for curvilinear coordinates, the Cartesian \lstinline{TreeMesh} can
be used. Depending on the particular choice of discretization methods, the
Cartesian \lstinline{TreeMesh} can be \SI{10}{\percent} to \SI{25}{\percent}
more efficient than the curvilinear meshes in \trixi.

\section{Assessment of Julia for simulation-focused scientific computing}
\label{sec:assessment-of-julia}

Julia was designed from scratch to be suitable for numerical computing
\cite{bezanson2017julia}. Its usefulness has been demonstrated, \eg, for
GPU programming \cite{besard2018juliagpu, omlin2020solving},
optimization \cite{dunning2017jump}, data science including
big data\footnote{Celeste project: \url{https://github.com/jeff-regier/Celeste.jl},
see \url{https://juliacomputing.com/case-studies/celeste} (accessed 2021-08-11)},
machine learning \cite{innes2018fashionable}, and scientific machine learning
(SciML) \cite{pal2021opening}. There are also mature libraries for ODE problems
\cite{rackauckas2017differentialequations}, small scale spectral approximations
\cite{olver2014practical}, and classical finite element methods \cite{badia2020gridap}.
In addition, there are some packages dedicated to solving specific time-dependent
PDEs such as \cite{ramadhan2020oceananigans, constantinou2021geophysicalflows}.
However, there does not seem to be a general framework for high-order methods
for hyperbolic PDEs in Julia. Thus, \trixi is an ideal candidate for a case study
of Julia for simulation-focused scientific computing, an area where codes written
in classical programming languages such as C and Fortran are still dominant
\cite{krais2021flexi, parsani2021ssdc}.

\subsection{What works well}

\subsubsection{Julia is fast}

As demonstrated in Section~\ref{sec:performance-comparison},
Julia is not generically slower than traditional high-performance programming
languages like C, C++, and Fortran. A minor exception in the context of scientific
computing relying on floating point operations is that Julia does not perform
automatic fused multiply-add (FMA) contraction, which can be remedied by
explicitly using the
\lstinline{muladd}\footnote{\url{https://docs.julialang.org/en/v1/base/math/\#Base.muladd}}
function or the more convenient
\lstinline{@muladd}\footnote{\url{https://github.com/SciML/MuladdMacro.jl}}
macro (for more details see the appendix).

\subsubsection{Julia encourages good software development practices}

Many scientists implementing numerical methods have no formal training
in software development. Julia and its ecosystem support these
researchers by making it easy to set up unit and regression tests, since a
testing framework is included in the standard library. This facilitates test-driven
development and continuous integration (CI), which makes code restructuring
and optimization straightforward.

Julia itself and most packages are developed on GitHub. The Julia community provides
several tools to run corresponding CI setups, prepare and publish
documentation\footnote{\url{https://github.com/JuliaDocs/Documenter.jl}},
and other related tasks. This setup also encourages other good software development
practices that we use in \trixi, such as mandatory code reviews.

In addition, the Julia package manager \lstinline{Pkg} fosters a unified form of semantic versioning
across the Julia ecosystem, which lays the basis for the following three
observations.

\subsubsection{It is easy to set up reproducible numerical experiments}

The package manager \lstinline{Pkg} makes it easy to reproduce the exact runtime environment, including
binary dependencies, used to generate numerical results. We have used this feature for all of our papers based on
\trixi \cite{schlottkelakemper2021purely, ranocha2021preventing}, including the present manuscript
\cite{ranocha2021adaptiveRepro}.
This facilitates code sharing and reproducible research in computational science,
which is arguably important but not yet mainstream \cite{barnes2010publish,
donoho2010invitation, leveque2013top}.

\subsubsection{External libraries can be integrated with relative ease}
\label{sec:external-libraries}

To incorporate some of the adaptive mesh capabilities into \trixi
we have created a Julia wrapper\footnote{\url{https://github.com/trixi-framework/P4est.jl}}
for the C library \texttt{p4est} \cite{burstedde2011p4est}. In our experience,
it is relatively easy to do so in Julia, and the package manager in combination
with BinaryBuilder.jl makes it convenient to distribute the necessary binaries.
In particular, users do not have to compile libraries on their own system
and the binaries are available on all major platforms including Linux, macOS,
and Windows.

Similarly, there is no need to compile HDF5 on a user system due to the
availability of HDF5.jl\footnote{\url{https://github.com/JuliaIO/HDF5.jl}}.
There is still work to be done on how to improve the experience in combination
with MPI on HPC clusters, but some promising
approaches exist (see Section~\ref{sec:distributed-memory}).

It is also possible to wrap Fortran libraries/tools in Julia. For example, we
have created the wrappers KROME.jl\footnote{\url{https://github.com/trixi-framework/KROME.jl}}
for KROME, a package to embed chemistry in astrophysical simulations
\cite{grassi2014krome}, and HOHQMesh.jl\footnote{\url{https://github.com/trixi-framework/HOHQMesh.jl}}
for HOHQMesh\footnote{\url{https://github.com/trixi-framework/HOHQMesh}}, a high order
hex-quad mesh generator written in Fortran.

\subsubsection{Packages can be used together at low (or no) cost}

Due to Julia's package manager, the cost of using external dependencies is low
enough to not become an issue. Traditionally, many scientific simulation codes tend
to reduce the number of external dependencies as much as possible due to the
complexity of handling different versions and making everything work together.
Such an approach often leads to significant code duplication. For example, many
CFD codes implement their own time integration methods, arguing that the spatial
part is much more complex. While this is often true, it is still more efficient
in terms of developer time to reuse existing implementations. This
allows experts on time integration methods to develop specialized algorithms and
implement them in open source software, while practitioners or researchers focusing
on spatial semidiscretizations can benefit with near-zero effort.
In our case, optimized time integration methods were developed in
\cite{ranocha2021optimized} and implemented in OrdinaryDiffEq.jl. These methods
can be used with \trixi by changing a single line of code. In contrast, researchers
working without external dependencies would need extra time to digest a new method's
details before even beginning an implementation. This additional overhead
makes it less likely that researchers would use these novel schemes in their own
codes and benefit from recent algorithmic developments.

An additional example is given by automatic differentiation. Due to multiple dispatch,
it is possible to conveniently use forward mode automatic differentiation
\cite{revels2016forward} to compute Jacobians of semidiscretizations of nonlinear
conservation laws with \trixi or to differentiate through a complete simulation,
including time integration methods from OrdinaryDiffEq.jl.

\subsubsection{Julia solves the two-language problem}

Many libraries for simulation-focused scientific computing are written in
languages such as C, C++, or Fortran to get decent performance. On top of these
low-level details, high-level wrappers are often provided to make it easier
for users to apply the algorithms for their problems. In contrast, \trixi is
written completely in Julia.

Such a difference between the programming languages used in the front end and
back end leads to a well-known barrier between a user and the developers of a library.
Julia was designed from the beginning to make it possible to write both simple
high-level code and highly efficient compute kernels, thereby solving the
two-language problem --- at least to a significant extent. Highly efficient low-level
code will still look different from simple high-level code, but the
barrier between them is much smaller. In particular, this makes it easier for
users to transition to developers and contribute to a package. This also contributes
to the following observation:

\subsubsection{The code base is simple enough to be useful for new users}

Due to the package manager, \trixi and all related ODE and visualization tools
can be installed with a single command, reducing the overhead of exploring the
code. In addition, Julia's expressiveness and the ability to work with a restricted
subset of all possible features have enabled already more than 18 students to use \trixi
for their coursework or theses. In our experience, the effort to get
started with traditional monolithic code bases is often so large that due to time restrictions
students either choose to build their own implementations specialized on their tasks or only work on
projects with a limited scope.

Another anecdotal example for the ease of use is the preprint \cite{singh2021linear}.
While the authors utilized \trixi for some numerical experiments, they did so
independently of the development team (who learned that the authors were
using \trixi only \textit{after} reading the preprint).

\subsubsection{Existing features can be extended and combined easily}

Due to Julia's high-level programming approach, dynamic typing, and multiple
dispatch, it is easy to combine existing functionality efficiently. For example,
the single-physics solvers for hyperbolic PDEs of \trixi were extended to
a multi-physics setup for the compressible Euler equations with self-gravity
with roughly 350 lines of code \cite{schlottkelakemper2021purely}.
In addition, \trixi can be extended from the outside without modifying the main
source code, which makes it easy to set up new simulation approaches and analyze
existing ones, \eg, by fluctuation simulations \cite{ranocha2021preventing}.

\subsubsection{Julia is free}

Julia itself is released under the MIT license and ships some GPL-licensed
third-party software (that can optionally be disabled for commercial purposes).
Many packages in the Julia ecosystem follow this approach and are freely available
under the permissive MIT license, including \trixi. This allows programmers to use Julia
without needing to pay for commercial software. In particular, students can
work with the software on their private computers without restrictions.

\subsection{What is still difficult or unknown}

\subsubsection{Compilation times can be annoying}

Since Julia uses a serial just-ahead-of-time compiler that currently does not cache
code between different Julia sessions, the initial compilation time can be
annoying. For example, the time to finish a first \trixi simulation in a fresh Julia
session and plot the results on a notebook can take between 30 seconds and a
minute. Having compiled the code, the second simulation and plot take less than
0.05 seconds. Thus, switching to Julia requires adapting the workflow compared
to other languages such as keeping a REPL session active for a longer time and
using packages such as Revise.jl\footnote{\url{https://github.com/timholy/Revise.jl}}.
This is particularly problematic in HPC environments. There are tools such as
PackageCompiler.jl\footnote{\url{https://github.com/JuliaLang/PackageCompiler.jl}}
that can help with these tasks but a good workflow usable for development and
deployment still has to be found.

\subsubsection{Simulations with distributed-memory parallelism}
\label{sec:distributed-memory}

While Julia offers its own approach to distributed computing via the \lstinline{Distributed} package
in the standard library, its performance and scalability are currently limited \cite{byrne2021mpi}.
As an alternative, the package MPI.jl utilizes the Message Passing Interface (MPI) for data
exchange \cite{byrne2021mpi}. It provides an API similar to MPI's C interface and allows one to either
wrap MPI binaries installed through Julia's package manager or to make use of an existing MPI
installation.  Most Julia projects focusing on distributed-memory simulations seem to be using
MPI.jl, such as ClimateMachine.jl\footnote{\url{https://github.com/CliMA/ClimateMachine.jl}} or
ImplicitGlobalGrid.jl\footnote{\url{https://github.com/eth-cscs/ImplicitGlobalGrid.jl}}.  In
general, the performance of MPI programs written in Julia is comparable to C programs
\cite{hunold2020benchmarking}, and parallel simulations with Julia have been shown to scale
to more than 5{,}000 GPUs with MPI.jl \cite{omlin2020solving}.

However, some practical challenges associated with running MPI-based parallel
simulations with Julia remain. For starters, the issue of compilation times, described in the previous
section, is exacerbated as MPI runs are by default not interactive and compilation results cannot be cached between MPI sessions. As such, this
requires code to recompile for each subsequent execution of an MPI-parallel Julia program,
which is particularly annoying during development. Partial relief can be obtained by using
\texttt{tmpi}\footnote{\url{https://github.com/Azrael3000/tmpi}} that allows the user to simultaneously
interact with multiple interactive Julia REPL sessions via the terminal multiplexer \texttt{tmux}.
However, this approach does not scale beyond a few MPI ranks and is only available on
Unix-like operating systems. The use of PackageCompiler.jl to reduce compilation times as much as
possible thus becomes mandatory. Another issue encountered when running Julia with many parallel
processes is that Julia operates with many small files during startup and precompilation,
putting heavy pressure on the parallel file system. This can be partially alleviated by doing
precompilation in serial. There are also tools like
\texttt{Spindle}\footnote{\url{https://github.com/hpc/Spindle}} that help solve similar
problems that occur when running Python programs in parallel \cite{frings2013parallelloading}.
Finally, the ease of using external libraries as described in Section~\ref{sec:external-libraries} is
somewhat lost when working with dependencies that themselves are parallelized with MPI. In this
case it is necessary to manually specify the paths to locally compiled libraries that have been
built against the respective system MPI installation, since there is generally no binary
compatibility between different MPI implementations. There are ongoing efforts in the Julia HPC
community to work around this limitation by using the
WI4MPI\footnote{\url{https://github.com/cea-hpc/wi4mpi}} wrapper package or the
Spack package manager \cite{gamblin2015spack,byrne2021mpi}.

Even though the aforementioned difficulties do not prevent massively parallel simulations with
Julia, they raise the entry barrier for new users and developers. The overarching issue, however, is
that at the time of writing, there is little precedence in terms of openly available tutorials,
citable publications, or highly scalable example applications that can be used to learn about best
practices for MPI-based parallelism in Julia. Thus, while in general it is justified to be cautiously
optimistic that most issues can be overcome, it is too early to make a final assessment of the
suitability of Julia for highly parallel simulations.

\section{Summary and conclusions}
\label{sec:summary}

We have presented \trixi, a Julia package for adaptive high-order numerical
simulations of hyperbolic PDEs. As researchers in numerical analysis and
scientific computing, our goals were to create a framework that is extensible,
easy to use, and fast (in this particular order). Making use of Julia's strengths,
we have been successful based on recent publications making use of \trixi,
the number of students and researchers working with \trixi, and serial performance
comparisons with a mature Fortran code. Having developed, from scratch, \trixi for a bit
more than one year allows us to give an assessment of Julia for simulation-focused
scientific computing.

Based on our experience, we consider Julia to be suitable for simulation-focused
scientific computing, in particular for hyperbolic PDEs, computational fluid
dynamics, and related problems --- at least on the scale of shared memory
parallelism. The scalability of high-order methods for hyperbolic PDEs written
in Julia to high-performance computing applications still needs to be demonstrated.
Yet, we do not consider this an unsolvable problem, since Julia is not
generically slower than traditional compiled HPC programming languages. Nevertheless,
it appears to be more complicated to scale Julia to distributed systems without losing some of the
simplicity and flexibility it offers for serial or shared-memory parallel computations.

Learning a new programming language naturally requires some time and effort.
In our experience, however, it has paid off and we do not want to miss the
many useful features of \trixi enabled by Julia and its ecosystem. While there is
no need to switch to Julia if people are satisfied with their existing
tools, we encourage researchers to try out Julia for scientific computing and to stay for a while.
The Julia
community (on the Julia Discourse forum\footnote{\url{https://discourse.julialang.org}}
or Slack/Zulip workspaces) is usually welcoming and helpful, both for new and
experienced users.

\section*{Acknowledgments}

We would like to thank all other contributors to and users of \trixi for their
help, encouragement, and discussions.
Funded by the Deutsche Forschungsgemeinschaft (DFG, German Research Foundation)
under Germany's Excellence Strategy EXC 2044-390685587, Mathematics Münster:
Dynamics-Geometry-Structure.
Andrew Winters was funded through Vetenskapsr{\aa}det, Sweden grant
agreement 2020-03642 VR.
Jesse Chan acknowledges support from the US National Science Foundation under
awards DMS-1719818 and DMS-1943186.
Gregor J. Gassner and Michael Schlottke-Lakemper received funding from the European Research Council through the ERC Starting Grant “An Exascale aware and Un-crashable Space-Time-Adaptive Discontinuous Spectral Element Solver for Non-Linear Conservation Laws” (Extreme), ERC grant agreement no. 714487.
The performance computations were enabled by resources provided by the Swedish National
Infrastructure for Computing (SNIC) at Tetralith partially funded by the Swedish
Research Council through grant agreement no. 2018-05973.


\bibliographystyle{juliacon}
\bibliography{ref.bib}

\appendix

Julia does not perform automatic fused multiply-add (FMA) contraction, i.e.,
replacing a multiplication followed by an addition with a single FMA instruction,
as it is inconsistent with its strict floating point semantics.
In other compilers, this optimization can be controlled by a global compiler option
(\lstinline{-ffp-contract} in GCC and Clang) and is often enabled at reasonable
optimization levels such as \lstinline{-O2} for GCC and Intel compilers\footnote{%
An interesting coincidence is that not only Julia but most LLVM-based frontends
tend not to enable FMA contraction by default.}.
In Julia, FMA contraction is controlled more locally: the
\lstinline{muladd} 
intrinsic function allows the compiler to evaluate a multiply-add using the most
efficient method. Relying on Julia's code manipulation techniques, the
\lstinline{@muladd} 
macro makes it convenient to use the \lstinline{muladd} function by
syntactically rewriting expressions to insert \lstinline{muladd} in appropriate
locations.

\end{document}